\documentclass[twocolumn,preprintnumbers,amsmath,amssymb,floatfix]{revtex4}

\usepackage{graphicx}% Include figure files
\usepackage{dcolumn}% Align table columns on decimal point
\usepackage{bm}% bold math

\begin{document}

\title{Dispersion management using betatron resonances in an ultracold-atom storage ring}

\author{K.W.\ Murch, K.L.\ Moore, S.\ Gupta, and D.M.\ Stamper-Kurn}

%\maketitle

\newcommand{\comment}[1]{}
\newcommand{\bdp}{B^{\prime \prime}_z}
\newcommand{\bp}{B^\prime}
\newcommand{\bs}{B_s}
\newcommand{\brot}{B_{\mbox{\scriptsize rot}}}
\newcommand{\omegares}{\Omega_{\mbox{\scriptsize res}}}
\newcommand{\bdpeff}{B^{\prime \prime}_{\mbox{\scriptsize eff}}}
\newcommand{\omegarot}{\omega_{rot}}
\newcommand{\deltar}{\delta r}
\newcommand{\deltaz}{\delta z}

\affiliation{Department of Physics, University of California,
Berkeley CA 94720}

\begin{abstract}
Specific velocities of particles circulating in a storage ring can
lead to betatron resonances at which static perturbations of the
particles' orbit yield large transverse (betatron) oscillations.
We have observed betatron resonances in an ultracold-atom storage
ring by direct observation of betatron motion. These resonances
caused a near-elimination of the longitudinal dispersion of atomic
beams propagating at resonant velocities, an effect which can
improve the performance of atom interferometric devices. Both the
resonant velocities and the strength of the resonances were varied
by deliberate modifications to the storage ring.
\end{abstract}

\maketitle

In circular accelerators and storage rings, the transverse
oscillations of guided particles away from the unperturbed beam
path are known, for historical reasons \cite{kers41}, as betatron
oscillations. While such oscillations can usually be kept small in
a well-designed device, large-scale betatron motion can
nevertheless be  excited resonantly by weak, static perturbations
of the beam path if the particle beam propagates at specific
velocities. In high energy devices \cite{cern94} and in neutron
storage rings \cite{kugl85}, such betatron resonances cause the
particle beam to collide with apertures in the device and be lost.

Today, storage rings, and, potentially, circular accelerators, are
being developed for ultracold, electrically neutral atoms
\cite{saue01,wu04guide,gupt05tort,arno05ring} or molecules
\cite{crom01ring}, to be used for precise interferometry using
guided matter waves \cite{wang05inter}, studies of low-energy
collisions \cite{bugg04dwave,thom04partial}, and the manipulation
of quantum degenerate matter.  That these motivations are quite
different from those for high-energy physics suggests that the
familiar concepts of accelerator physics be examined in a new
light.

Here, we explore betatron resonances in a millimeter-scale storage
ring for ultracold atoms with energies of around $k_B \times 100
\, \mu$K per atom (100 peV per nucleon), orders of magnitude
smaller than those in modern particle physics facilities. The
exchange of energy between longitudinal and transverse motion at
several betatron resonances is directly observed for Bose-Einstein
condensates (BECs) propagating in the ring. This exchange
dramatically reduces the longitudinal dispersion of the atomic
beam, lowering its longitudinal kinetic temperature to the pK
range. Such dispersion management may greatly improve measurements
of rotation rates \cite{reih91rot,lene97,gust97}, fundamental
constants \cite{weis93}, and other quantities \cite{berm96int} by
atom-interferometric schemes which are sensitive only to
longitudinal velocities. Our study emphasizes the conceptual unity
among storage rings,  and suggests that other concepts of
accelerator physics, such as non-linear resonant beam extraction
or Landau damping of modulation instabilities, may also be adapted
to the ultralow energy domain.

\begin{figure}
\includegraphics[angle = 0, width = 0.4\textwidth]{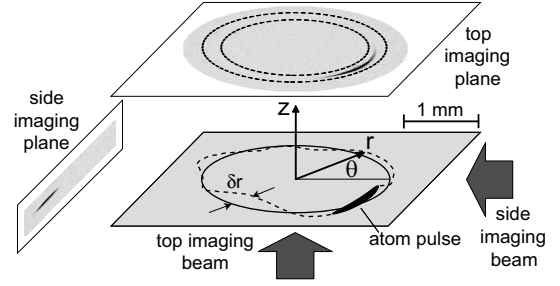}
\caption{An ultracold-atom storage ring. Axes are indicated with
gravity along $-\hat{z}$.  Top or side view images of the atoms
allow their motion to be studied.  Dashed circles indicate the 270
$\mu$m radial extent of top-view images displayed in polar
coordinates as ``annular views'' in Figs.\ \ref{fig:disp} and
\ref{fig:ressys}. We refer to motion in $r$ as radial, in $z$ as
axial, and in $\theta$ as azimuthal or longitudinal. The nominally
circular storage ring contains radial $\deltar$ (exaggerated for
illustration) and axial $\deltaz$ beam-path errors and an
azimuthal variation $U(\theta)$ in the
potential.\label{fig:setup}}
\end{figure}

The storage ring used for this work is a magnetic time-orbiting
ring trap for ultracold $^{87}$Rb atoms \cite{gupt05tort} (Fig.\
\ref{fig:setup}). Atoms are confined to a circular ring of radius
$r = 1.25$ mm in the horizontal plane by a transverse potential
which is harmonic for small displacements from the nominal beam
orbit. Both frequencies for transverse motion in this potential,
i.e.\ the betatron frequencies for radial ($\omega_r$) and for
axial ($\omega_z$) motion, are independently adjusted by varying
the time-varying magnetic field $\brot$ used in our trap
\cite{brotfootnote}.  For atoms in the $|F=1, m_F=-1\rangle$
hyperfine ground state used in this work, using $\brot$ between
4.8 and 13$\,$G gives betatron frequencies in the range $2
\pi\,(50 - 90)\,$Hz. This range of $\brot$ varies the maximum
height of the trapping potential in the transverse directions in
the range of $k_B\,(40 - 100)\,\mu$K.

%Figure 1

Pulsed atomic beams are ``injected'' into the storage ring in two
stages. First, a non-degenerate gas of $2 \times 10^7$ atoms is
loaded in a portion of the storage ring, with an azimuthal
confining potential added to the ring by application of a 9 G
sideways magnetic field \cite{gupt05tort}. This gas is then cooled
using forced evaporation to produce a BEC of up to $3 \times 10^5$
atoms. Second, the BEC is accelerated over $30\,$ms by changing
the sideways field so as to lower the potential of the trap
minimum and advance its position along the ring.  By varying
settings in this acceleration, the mean initial angular velocity
of atoms in the ring, $\Omega_{i}$, is varied between 40 and
120$\,$rad/s \cite{measurementfootnote}. The sideways field is
eliminated over the next $30\,$ms as $\brot$ is brought to its
storage-ring setting.

%check speeds for non-uniform acceleration

Important parameters for characterizing betatron motion in a
storage ring are the tune parameters, $\nu_{r,z} = \omega_{r,z} /
\Omega$, which give the number of betatron oscillations occurring
over one complete orbit along the closed-loop beam path.  At
integer values of the tune parameters, the transverse deflections
of particles due to small, spatially-fixed perturbations to the
beam path accumulate in phase over many orbits to drive large
scale betatron motion.  This describes a low-order betatron
resonance which is the subject of the present work.  Higher-order
perturbations, such as a spatial modulation of the transverse
trapping frequencies or coupling between the two modes of betatron
motion, can result in higher-order betatron resonances at
non-integer values of the tune parameters and in non-linear
coupling between resonances \cite{cern94}.  We did not observe
these higher-order effects in our storage ring.

\begin{figure}
\includegraphics[angle = 0, width = 0.4\textwidth] {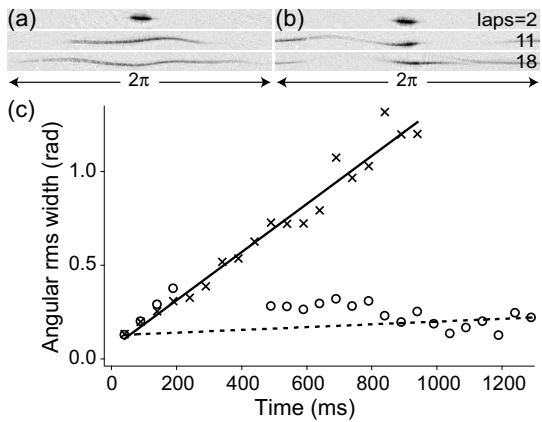}
\caption{Dispersion management of matter waves in a storage ring.
Annular (top) views are shown after 2, 11, and 18 complete
revolutions for initial mean axial tunes of (a) $\nu_z = 4.2$ and
(b) $\nu_z = 4.0$.  Slight bends in the cloud result from radial
betatron motion excited during the injection sequence. (c) The rms
width of all (X's for $\nu_z = 4.2$) or just the compact portion
(circles for $\nu_z = 4.0$) of the atomic beam is shown vs.\
propagation time, with data for $\nu_z=4.0$ limited to times when
the compact and diffuse portions are separated. A linear fit to
data for $\nu_z=4.2$ (solid line), and a line joining earliest and
latest data for $\nu_z = 4.0$ (dashed line) are shown. Here
$(\omega_r,\omega_z) = 2 \pi\,(48,60)$ Hz.\label{fig:disp}}
\end{figure}

The propagation of an atomic beam launched at two different
initial velocities, either away from or at a betatron resonance,
is compared in Fig.\ \ref{fig:disp}. At the measured axial tune of
$\nu_z = 4.2$, the rms azimuthal width of the cloud grows steadily
according to an rms azimuthal linear velocity of $\Delta v =
1.7\,$mm/s \cite{velfootnote}, equivalent to an rms variation in
the tune of $\Delta \nu_z = 6 \times 10^{-2}$, or to a
longitudinal kinetic temperature of $T = m (\Delta v)^2 / k_B =
28$ nK. At this rate of expansion, the propagating cloud fills a
large portion of the storage ring within 10 revolutions.

%Figure 2

An atomic beam launched at a resonant tune parameter evolves quite
differently. With the mean tune parameter $\nu_z = 4.0$ set to an
integer value, about half of the atoms are ``caught'' in a
betatron resonance, and their longitudinal dispersion ceases so
that  a compact portion of the beam remains even after 18
revolutions (Fig.\ \ref{fig:disp}b). Starting after about
$250\,$ms, the width of this compact portion shows no further
growth. Based on measurements from the earliest and latest times
in Fig.\ \ref{fig:disp}c, we estimate the rms velocity of this
portion of the beam to be below $100\,\mu$m/s, reducing the range
of tunes to $\Delta \nu_z$\,$<$\,$4 \times 10^{-3}$. This
corresponds to a longitudinal kinetic temperature of below
$100\,$pK. To our knowledge, this is the lowest kinetic
temperature ever reported for an atomic beam.

%Figure 3

\begin{figure}
\includegraphics[angle = 0, width = 0.4\textwidth] {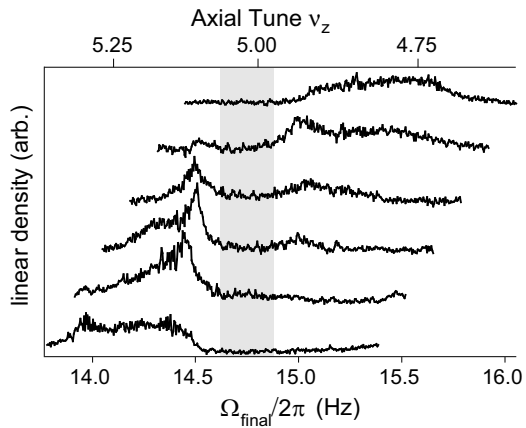}
\caption{Matter wave dispersion at an axial betatron resonance.
The distribution of final azimuthal velocities is shown for beams
with initial mean angular velocities $\Omega_i / 2 \pi$ evenly
spaced between 14.3 (bottom curve) and 15.4 Hz (top curve). Data
are offset vertically for clarity. These distributions are
obtained from the radially-integrated column density in top view
images taken after 640 ms of propagation. Atoms expelled from the
stopband (gray shading) accumulate at its low-velocity edge. Since
the stopband is narrower than the full initial range of velocities
in the beam, only a portion of the beam is affected. Low-velocity
atoms for $\Omega_i/2 \pi = 14.3$ Hz are affected by the $\nu_r=5$
radial betatron resonance. Here $(\omega_r,\omega_z)=2
\pi\,(70,74)\,$Hz.\label{fig:resl5}}
\end{figure}

We note that the rms velocity of the atomic beam was not directly
measured, as could be done, for example, using Doppler-sensitive
spectroscopy \cite{sten99brag}.  Nevertheless, we argue that our
estimate for the rms velocity is indeed valid, especially in light
of the agreement with simulations of the resonance, as described
below. An alternative explanation for reduced dispersion due to
bright soliton formation \cite{stre02,khay02} appears ruled out
for two reasons. First, the interactions between $^{87}$Rb atoms
in our ring are repulsive, rather than attractive, while quantum
states with a negative effective mass, necessary for the creation
of gap solitons \cite{eier04}, do not appear to arise in our
storage-ring potential. Second, we also observe a reduced
longitudinal dispersion for atomic beams derived from a
non-degenerate atomic gas, albeit with greater dispersion than
when BECs are launched into the storage ring.  These beams have a
higher initial kinetic temperature and lower density than those
derived from a BEC, conditions which further disfavor soliton
formation.

The effects of the betatron resonance are displayed in more detail
in Fig.\ \ref{fig:resl5}.  Here, atomic wave packets are launched
at tune parameters near $\nu_z = 5$ and then imaged after 640 ms
of propagation.  The data indicate a range of velocities, also
known as a stopband \cite{cern94}, which cannot be maintained in
the static potential of the storage ring. While in a high energy
storage ring particles with velocities within the stopband are
lost, in our ring the transverse oscillations at the betatron
resonance have energies ($< k_B \times 5\, \mu$K) which are
insufficient to expel atoms from the ring. Rather, these atoms
remain in the storage ring and accumulate in a narrow range of
final velocities at the low-velocity edge of the stopband.

%Figure 4

\begin{figure}
\includegraphics[angle = 0, width = 0.4\textwidth] {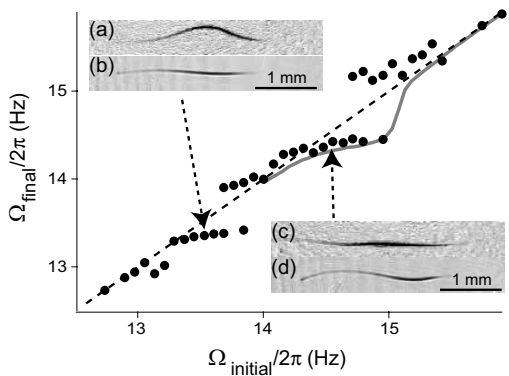}
\caption{Stopbands for $\nu_r=5$ radial (lower $\Omega_i$) and the
$\nu_z=5$ axial (higher $\Omega_i$) betatron resonances. Mean
final angular velocities, determined after 440 ms of propagation
in the ring, are shown vs.\ $\Omega_i$. Near the resonances, the
atomic beam divides into two portions (Fig.\ \ref{fig:resl5}), and
mean angular velocities of each are given. Simulation results for
the axial resonance (solid line), and the relation $\Omega_i =
\Omega_f$ (dashed line) are shown. The (a) annular and (b) side
views of the atoms launched at $\nu_r=5$ (radial resonance), and
(c) annular and (d) side views of those at $\nu_z=5$ (axial
resonance), show which transverse oscillation is enhanced at each
resonance. The horizontal scale for (a) and (c) is chosen to give
equal cloud lengths in annular- and side-view images. Here,
$(\omega_r,\omega_z)/2\pi = (69,73)\,$Hz.\label{fig:ressys}}
\end{figure}

Scanning the beam velocity over a wider range, several betatron
resonances can be observed, each yielding a stopband of disallowed
velocities (Fig.\ \ref{fig:ressys}). Resonances are identified as
either axial or radial by directly observing the betatron motion.
A radial resonance produces an atomic beam which oscillates
radially with a phase that differs across the length (i.e.\ across
azimuthal velocities) of the cloud. A side image of the same beam
shows no significant axial betatron motion. Conversely, an axial
betatron resonance induces axial but not radial oscillations. Such
images also confirm that the energy of the atomic beam is
conserved at a betatron resonance. For instance, the $60 \, \mu$m
maximum amplitude of radial betatron motion observed in Fig.\
\ref{fig:ressys}(a) corresponds to an energy of $k_B \times
3.4\,\mu$K. Given the initial kinetic energy of $k_B \times
62\,\mu$K for atoms launched at $\nu_r = 5$, this motion should
reduce the longitudinal velocity by $2.7\%$, in agreement with the
measured magnitude of the stopband.

The atomic beam in these experiments derives from a BEC and is
thus characterized by a macroscopic quantum wavefunction. Away
from a resonance, its transverse velocity width is consistent with
that of a minimum uncertainty quantum state, as determined from
its transverse expansion after a sudden release from the storage
ring. Nevertheless, since the energy transferred to betatron
motion is several hundred times larger than $\hbar \omega_{r,z}$,
the relevant harmonic oscillator energy quanta, a purely classical
treatment suffices to describe the atomic motion near a resonance.

%Figure 5

\begin{figure}
\includegraphics[angle = 0, width = 0.4\textwidth] {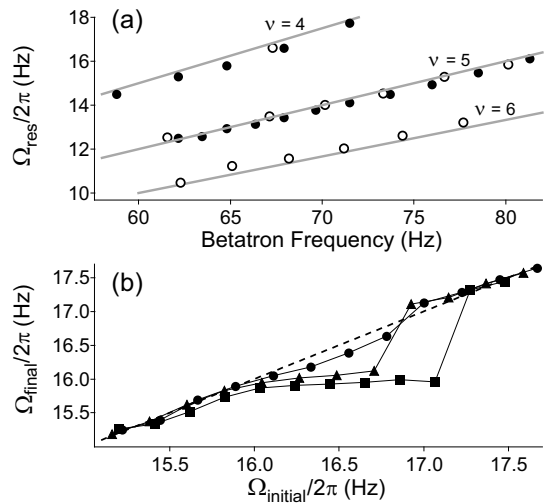}
\caption{Tuning the resonant velocities and strength of betatron
resonances. (a) Measured angular frequencies $\omegares$ at the
low velocity edge of stopbands observed at several radial (open
circles) and axial (filled circles) betatron resonances are shown
vs.\ measured betatron frequencies. (b) The $\nu_r=4$ resonance
for $(\omega_r,\omega_z)=2\pi\,(68,74)\,$Hz is characterized with
a deliberately applied $U(\theta) = U_4 \sin 4 \theta$ azimuthal
potential of strength $U_4/k_B=24$ (circles), $45$ (triangles),
and $81\,$nK (squares). The dashed line shows the relation
$\Omega_i = \Omega_f$.\label{fig:restune}}
\end{figure}

To model an axial betatron resonance, for example, we consider
that the locus of potential minima for our magnetic ring deviates
from the $z=0$ horizontal plane by an amount $\deltaz(\theta)$.
Such a deviation, or beam-path error, arises from imperfections in
the electromagnets used to generate the magnetic potential and
from stray magnetic fields.  Assuming $\deltaz$ is small, the
betatron resonance at each integer value of the axial tune arises
from beam-path errors with harmonic index $q = \nu_z$ in the
Fourier series $\deltaz(\theta) = \sum_{q=1}^\infty \deltaz_q
\sin(q \theta + \theta_q)$. The strength of the betatron
resonance, given, for example, by the magnitude of the stopband,
is then determined by the appropriate magnitude $\deltaz_q$.
Radial betatron resonances, which are caused by radial beam-path
errors $\deltar(\theta)$ and azimuthal variations in the potential
minimum $U(\theta)$, can be similarly analyzed.

For our simulation, we chose $\deltaz = \deltaz_5 \sin(5 \theta)$
while setting $\deltar$ and $U(\theta)$ to zero. The magnitude
$\deltaz_5 = 1.2 \mu$m was chosen to match the measured magnitude
of the resonance stopband. We then numerically integrated the
classical equations of motion for a point particle for $500\,$ms
with the initial condition that the particle propagate purely
azimuthally at angular frequency $\Omega_i$ with $z=0$ and radius
corresponding to a circular orbit in the absence of beam-path
errors. The simulations indicate that for $\omega_z / \Omega_i
\simeq 5$, a fraction of the total energy is exchanged between the
azimuthal and axial motions, reducing the average longitudinal
velocity. The simulations confirm that particles initially
propagating at tunes within the stopband are slowed to a narrow
range of average velocities, the narrowness of which accounts for
the reduced dispersion at the betatron resonance.

The manipulation of matter waves through betatron resonances
represents a novel approach to manage dispersion in a circular,
multimode atomic waveguide. To make use of such dispersion
management (see also Ref. \cite{eier03}), it is desirable to
control both the resonant velocity and also the strength of a
betatron resonance. We demonstrate both these capabilities in our
storage ring.  To vary the betatron resonance frequency, and
thereby the resonance velocity, we adjust the transverse trap
frequencies in our storage ring.  The measured positions of
several resonances at various settings of our storage ring, shown
in Fig.\ \ref{fig:restune}a, exhibit the expected linear scaling
of resonant angular velocities with the betatron frequencies.  To
vary the strength of a specific resonance, we add to our magnetic
trap a static radial-quadrupole magnetic field with its axis
coinciding with that of the storage ring. This magnetic field adds
two types of errors to the ring: a $q=2$ axial beam-path error,
and a $q=4$ potential error $U(\theta)$ which is proportional to
the azimuthal field magnitude. As shown in Fig.\
\ref{fig:restune}(b), the strength of the radial betatron
resonance at $\nu_r=4$, assessed by the magnitude of the stopband,
is adjusted by varying the magnitude of the deliberately applied
$q=4$ modulation.

Our work has important implications for guided-atom
interferometry. For example, an atom beam produced at an axial
betatron resonance will experience less variation in the Coriolis
acceleration produced in a frame rotating about the ring axis, as
a result of having a narrowly defined orbital radius and angular
velocity. An atomic Sagnac interferometer \cite{gust97} could thus
detect rotations with higher precision. On the other hand, the
excitation of betatron motion \cite{lean02guide,brom04bends} may
add uncertain path-dependent phases in an atom interferometer. The
effects of betatron resonances, which cause even small defects in
a waveguide to induce large betatron oscillations of guided atoms,
will require further scrutiny as guided-atom interferometers are
developed and deployed.

We thank J.\ Wurtele, D.\ Meschede, and W.\ Ketterle  for helpful
comments. This work was supported by DARPA (Contract No.\
F30602-01-2-0524), and the David and Lucile Packard Foundation.
KLM acknowledges support from NSF, and SG from the Miller
Institute.

%\bibliographystyle{prsty}
%\bibliography{dcrefs10_dmsk,allrefs}

\begin{thebibliography}{10}

\bibitem{kers41}
D.~W. Kerst and R. Serber, Phys. Rev. {\bf 60},  53  (1941).

\bibitem{cern94}
{\em CAS - CERN Acclerator School: 5th General Accelerator Physics
Course},
  edited by S. Turner (CERN, Geneva, Switzerland, 1994).

\bibitem{kugl85}
K.-J. K{\"u}gler, K. Moritz, W. Paul, and U. Trinks, Nucl. Instrm.
Methods {\bf
  228},  240  (1985).

\bibitem{saue01}
J.~A. Sauer, M.~D. Barrett, and M.~S. Chapman, Phys. Rev. Lett.
{\bf 87},
  270401  (2001).

\bibitem{wu04guide}
S. Wu, W. Rooijakkers, P. Striehl, and M. Prentiss, Phys. Rev. A
{\bf 70},
  013409  (2004).

\bibitem{gupt05tort}
S. Gupta {\it et~al.}, preprint, arXiv:cond-mat/0504749.

\bibitem{arno05ring}
A.~S. Arnold, C.~S. Garvie, and E. Riis, preprint,
arXiv:cond-mat/0506142.

\bibitem{crom01ring}
F.~M.~H. Crompvoets, H.~L. Bethlem, R.~T. Jongma, and G. Meijer,
Nature {\bf
  411},  174  (2001).

\bibitem{wang05inter}
Y.-J. Wang {\it et~al.}, Phys. Rev. Lett. {\bf 94},  090405
(2005).

\bibitem{bugg04dwave}
C. Buggle, J. Leonard, W.~von Klitzing, and J.~T.~M. Walraven,
Phys. Rev. Lett.
  {\bf 93},  173202  (2004).

\bibitem{thom04partial}
N.~R. Thomas, N. Kjaergaard, P.~S. Julienne, and A.~C. Wilson,
Phys. Rev. Lett.
  {\bf 93},  173201  (2004).

\bibitem{reih91rot}
F. Riehle {\it et~al.}, Phys. Rev. Lett. {\bf 67},  177  (1991).

\bibitem{lene97}
A. Lenef {\it et~al.}, Phys. Rev. Lett. {\bf 78},  760  (1997).

\bibitem{gust97}
T.~L. Gustavson, P. Bouyer, and M.~A. Kasevich, Phys. Rev. Lett.
{\bf 78},
  2046  (1997).

\bibitem{weis93}
D.~S. Weiss, B.~C. Young, and S. Chu, Phys. Rev. Lett. {\bf 70},
2706  (1993).

\bibitem{berm96int}
P.~R. Berman, {\em Atom Interferometry} (Academic Press, New York,
1996).

\bibitem{brotfootnote}
The time-varying field, evaluated at a point on the ring, is $B_r
  \sin(\omegarot t) \hat{r} + B_z \cos(\omegarot t)\hat{z}$, with $\omegarot =
  2 \pi \times 5$ kHz. Independent control over $\omega_{r,z}$ is obtained by
  varying $B_r$ and $B_z$ separately. We use the time-averaged magnitude
  $\brot$ as shorthand. The static transverse gradient at the ring is $\bp =
  310$ G/cm. See Ref.\ \cite{gupt05tort} for details.

\bibitem{measurementfootnote}
A linear relation between $\Omega_i$ and the sideways field used
to accelerate
  the atoms is obtained by measuring beam velocities at several settings far
  from betatron resonances, and then used to determine $\Omega_i$ at all
  settings. Betatron frequencies are determined by exciting and then observing
  betatron motion of the propagating atomic beam.

\bibitem{velfootnote}
We relate azimuthal linear ($v$) and angular ($\Omega$) velocities
by $v =
  \Omega \, r(\Omega)$. The beam-path radius $r(\Omega)$ depends on $\Omega$
  due to centrifugal effects.

\bibitem{sten99brag}
J. Stenger {\it et~al.}, Phys. Rev. Lett. {\bf 82},  4569  (1999).

\bibitem{stre02}
K.~E. Strecker, G.~B. Partridge, A.~G. Truscott, and R.~G. Hulet,
Nature {\bf
  417},  150  (2002).

\bibitem{khay02}
L. Khaykovich {\it et~al.}, Science {\bf 296},  1290  (2002).

\bibitem{eier04}
B. Eiermann {\it et~al.}, Phys. Rev. Lett. {\bf 92},  230401
(2004).

\bibitem{eier03}
B. Eiermann {\it et~al.}, Phys. Rev. Lett. {\bf 91},  060402
(2003).

\bibitem{lean02guide}
A.~E. Leanhardt {\it et~al.}, Phys. Rev. Lett. {\bf 89},  040401
(2002).

\bibitem{brom04bends}
M.~W.~J. Bromley and B.~D. Esry, Phys. Rev. A {\bf 69},  053620
(2004).

\end{thebibliography}

\end{document}